\documentclass{article}
\usepackage{isabelle,isabellesym}
\usepackage{hyperref}
\usepackage[numbers]{natbib}

\usepackage{amsmath}
\usepackage{amsthm}
\usepackage{amsfonts}
\usepackage{amssymb}
\usepackage{bbm}  %
\usepackage{enumitem}
\usepackage{multidef}
\usepackage{verbatim}
\usepackage{stmaryrd} %
\renewcommand{\o}{\vee}

\newcommand{\y}{\wedge}

\newcommand{\limp}{\longrightarrow}
\newcommand{\lsii}{\longleftrightarrow}

\DeclareMathOperator{\dom}{dom}

\newcommand{\modelo}[1]{\mathbf{#1}}
\newcommand{\axiomas}[1]{\mathit{#1}}
\newcommand{\clase}[1]{\mathsf{#1}}

\newcommand{\operador}[1]{\mathbf{#1}}

\multidef{\clase{#1}}{Card,HC,HF,Lim,On->Ord,Reg,WF,Ord}

\DeclareMathAlphabet{\mathbbm}{U}{bbm}{m}{n} 
\newcommand{\1}{\mathbbm{1}}
\newcommand{\PP}{\mathbbm{P}}

\multidef[prefix=cal]{\mathcal{#1}}{A-Z}
\multidef{\modelo{#1}}{A,BB->B,CC->C,NN->N,QQ->Q,RR->R,ZZ->Z}

\multidef[prefix=p]{\mathbb{#1}}{A-Z}

\renewcommand{\H}{\operador{H}}
\renewcommand{\S}{\operador{S}}

\renewcommand{\emptyset}{\varnothing}

\newcommand{\Pow}{\mathop{\mathcal{P}}}
\renewcommand{\P}{\Pow}

\newcommand{\R}{\mathbb{R}}
\newcommand{\N}{\mathbb{N}}

\newcommand{\lb}{\langle}
\newcommand{\rb}{\rangle}

\renewcommand{\phi}{\varphi}

\newcommand{\defi}{\mathrel{\mathop:}=}

\newcommand{\om}{\ensuremath{\omega}}

\newcommand{\AC}{\axiomas{AC}}
\newcommand{\DC}{\axiomas{DC}}

\newcommand{\CH}{\axiomas{CH}}
\newcommand{\ZFC}{\axiomas{ZFC}}

\newcommand{\rr}{\mathrel{R}}

\newcommand{\sm}{\setminus}
\newcommand{\sbq}{\subseteq}

\newcommand{\mty}{\emptyset}

\DeclareMathOperator{\val}{\mathit{val}}
\DeclareMathOperator{\chk}{\mathit{check}}

\makeatletter
\@ifpackageloaded{txfonts}\@tempswafalse\@tempswatrue
\if@tempswa
  \DeclareFontFamily{U}{txsymbols}{}
  \DeclareFontFamily{U}{txAMSb}{}
  \DeclareSymbolFont{txsymbols}{OMS}{txsy}{m}{n}
  \SetSymbolFont{txsymbols}{bold}{OMS}{txsy}{bx}{n}
  \DeclareFontSubstitution{OMS}{txsy}{m}{n}
  \DeclareSymbolFont{txAMSb}{U}{txsyb}{m}{n}
  \SetSymbolFont{txAMSb}{bold}{U}{txsyb}{bx}{n}
  \DeclareFontSubstitution{U}{txsyb}{m}{n}
  \DeclareMathSymbol{\aleph}{\mathord}{txsymbols}{64}
  \DeclareMathSymbol{\beth}{\mathord}{txAMSb}{105}
  \DeclareMathSymbol{\gimel}{\mathord}{txAMSb}{106}
  \DeclareMathSymbol{\daleth}{\mathord}{txAMSb}{107}
\fi
\makeatother

\newtheorem{theorem}{Theorem}

\newtheorem*{claim*}{Claim}
\theoremstyle{definition}
\newtheorem{definition}[theorem]{Definition}

\theoremstyle{remark}
\newtheorem*{remark*}{Remark}

\makeatletter
\def\foottext{\gdef\@thefnmark{}\@footnotetext}
\makeatother
\newcommand{\keywords}[1]{\foottext{\emph{Keywords:} #1}}
\newcommand{\ack}[1]{\par\bigskip \noindent \emph{Acknowledgment:} #1}

\begin{document}
\title{First steps towards a formalization of Forcing}
\author{Emmanuel Gunther
  \and 
  Miguel Pagano
  \and 
  Pedro S\'anchez Terraf}
\maketitle

\begin{abstract} 
  We lay the ground for an Isabelle/ZF formalization of Cohen's technique of
  \emph{forcing}. We formalize the definition of forcing notions as
  preorders with top, dense subsets, and generic filters. We formalize
  a version of the principle of Dependent Choices and using it
  we prove the Rasiowa-Sikorski lemma on the existence of generic filters.
  
  Given a transitive set $M$, we define its generic extension $M[G]$,
  the canonical names for elements of $M$, and finally show that if $M$
  satisfies the axiom of pairing, then $M[G]$ also does. We also prove
  $M[G]$ is transitive.
\end{abstract}

\section{Introduction}\label{sec:introduction}
Set Theory plays a double role in Mathematics: It is one of its
possible foundations and also an active research area.
As it is widely known, Georg Cantor introduced its main concepts and in
particular showed the fundamental result that the real line, $\R$  is not
equipotent to the natural numbers. Soon after this, he posed the
most important question in the field, written as a conjecture:
\begin{quote}
  The \emph{Continuum Hypothesis} ($\CH$). Every uncountable subset of $\R$ is
  equipotent to $\R$.
\end{quote}

The current axiomatic foundation of Set Theory is through first-order
logic and uses the axioms devised by Zermelo and Fraenkel, including
the Axiom of Choice ($\AC$) among them. This theory is known by the
$\ZFC$ acronym. G\"odel \cite{godel-L} showed that $\CH$ cannot be refuted using
$\ZFC$, unless this theory itself is inconsistent (we say that
\emph{$\CH$ is relatively consistent with $\ZFC$}). For a while, this
result left the possibility that one might be able to show
$\ZFC\models \CH$, but in a groundbreaking work \cite{Cohen-CH-PNAS},
Paul Cohen discovered the technique of \emph{forcing} and proved that
$\neg\CH$ is relatively consistent with $\ZFC$. Forcing has been used
since then for showing innumerable independence results and to perform
mathematical constructions.

A great part of G\"odel's work on this subject has been formalized in
Isabelle~\cite{Isabelle} by Lawrence Paulson~\cite{paulson_2003}. This paper formalizes
a first part of the machinery of forcing, mostly by following the new
edition of the classical book on the subject by Kunen \cite{kunen2011set}. In the rest of the
introduction we discuss some of the set-theoretical details involved
and explain briefly Paulson's formalization.

\subsection{Models of $\ZFC$}
By G\"odel's Second Incompleteness Theorem, we cannot  prove that
there exists a model of $\ZFC$. More formally, if we assume that
mathematical proofs can be encoded as theorems of $\ZFC$ and that
the latter do not lead to contradictions (i.e., $\ZFC$ is
\emph{consistent}), then we cannot prove that there exists a set $M$
and a binary relation $E$ such that $\lb M,E\rb$ satisfies the $\ZFC$
axioms.

A relative consistency proof for an axiom $A$ is then obtained by
assuming that there exists a model of ZFC, say $\lb M,E\rb$, and
constructing another model $\lb M',E'\rb$ for $\ZFC + A$. We single
out a very special kind of models:
\begin{definition}\label{def:transitive-model}
  \begin{enumerate}
  \item A set $M$ (of sets) is \emph{transitive} if for all $x\in M$ and 
    $y\in x$, we have $y\in M$ (i.e., every element of $M$ is a subset
    of $M$).
  \item $\lb M,E\rb$ is a \emph{transitive model} if $M$ is transitive
     and  $E$ is the membership relation $\in$ restricted to
    $M$. It is \emph{countable} if $M$ is equipotent to a subset of
    $\N$; we then say that the model  $M$ is a \emph{ctm}. 
  \end{enumerate}
\end{definition}
\noindent As in the last sentence, one usually refers to a transitive
model by the underlying set because the relation is fixed.
In spite of G\"odel's Second Incompleteness Theorem, one can find
transitive models for every finite fragment of $\ZFC$. More precisely,
\begin{theorem}\label{th:ctm-finite-axioms}
  For each finite subset
  $\Phi\sbq \ZFC$, the statement \emph{``there exists a countable
    transitive model of $\Phi$''} is a theorem of $\ZFC$.
\end{theorem}
\noindent This follows by a combination of the Reflection Principle, the
L\"owenheim-Skolem Theorem, and the Mostowksi Collapse. The reader can
consult the details in \cite{kunen2011set}. Consistency arguments that
assume the existence of a ctm $M$ of $\ZFC$ can usually be replaced by
a model as in Theorem~\ref{th:ctm-finite-axioms}, since a first-order
proof (e.g.\ of a contradiction)%
\footnote{It is relevant to this point that 
  both the approaches by G\"odel and Cohen for showing
  relative consistency of an axiom $A$ 
  can be used to obtain an algorithm transforming a proof
  concluding a contradiction from $\ZFC+A$ to one from $\ZFC$.}
involves only finitely many axioms.

It is instructive to sketch G\"odel's argument of the relative
consistency of $\CH$: Assuming that $M$ is a ctm of $\ZFC$, G\"odel
showed that $M$ contains a minimal submodel $L^M$ of the same
``height'' (i.e.\ having the same ordinals) that satisfies
$\ZFC+\CH$. The sets in $L^M$ are called \emph{constructible} and are
in a sense ``definable.'' In fact, there is a first-order formula $L$
such that $L^M = \{x\in M : M\models L(x)\}$. To show that
$L^M\models \ZFC+\CH$, one uses the fact that $\ZFC$ holds in $M$.

It is therefore a primary need to have a means to correlate  (first-order)
properties satisfied by a model $M$ and those of a 
submodel $N\sbq M$. As a simple example on this, consider 
$M\defi \{a,b,c, \{a,b\},\{a,b,c\}\}$ and
$N\defi\{a,b,\{a,b,c\}\}$, and let 
\[
\phi(x,y,z)\defi \forall w.\,( w\in z \lsii w=x \o w=y).
\]
Then we have
\[
M\not\models \phi(a,b,\{a,b,c\}) \quad\text{ but }\quad N\models \phi(a,b,\{a,b,c\}).
\]
There is a discrepancy between  $M$ and $N$ about $\{a,b,c\}$ being ``the
(unordered) pair of $a$ and $b$.'' We say that $\phi$ holds for
$a,b,\{a,b,c\}$ \emph{relative} to $N$. It is immediate to see that
$\phi$ holds  for $x,y,z$ relative to $N$ if and only if
\[
\phi^N(x,y,z)\defi \forall w.\ w\in N\limp ( w\in z \lsii w=x \o w=y)
\] 
holds. $\phi^N$ is called the \emph{relativization of $\phi$ to
  $N$}. One can generalize this operation of relativization to the
class of all sets satisfying a first-order predicate $C$ in a
straightforward way:
\[
\phi^C(x,y,z)\defi \forall w.\ C(w)\limp ( w\in z \lsii w=x \o w=y)
\] 

It can be shown elementarily that if $M$ and $N$ are transitive,
$\phi^N$ holds if and only if $\phi^M$ holds,  for $x,y,z\in N$. We
say then that $\phi$ is \emph{absolute between $N$ and $M$.}
The concepts of relativization and absoluteness are central to the
task of transferring truth of axioms in $M$ to $L^M$, and constitute
the hardest part of Paulson's development.

\subsection{Forcing}
Forcing is a technique to extend countable transitive models of
$\ZFC$. This process is guaranteed to preserve the $\ZFC$
axioms while allowing to fine-tune what other first-order properties the
extension will have. Given a ctm $M$  of $\ZFC$ and a set $G$, one constructs a new
ctm  $M[G]$  that includes $M$ and
contains $G$, and proves that under some hypotheses ($G$ being ``generic''),
$M[G]$ satisfies $\ZFC$.

The easiest way to define genericity is by using a preorder with top
$\lb\PP,\leq,\1\rb$ in $M$.   
In Section~\ref{sec:forcing-notions} we formalize the definitions of
\emph{dense} subset and  \emph{filter} of  $\PP$, and we say that  $G$
is an $M$-generic filter
if it intersects every dense subset of $\PP$ that lies in $M$.

The Rasiowa-Sikorski lemma (RSL) states that for any preorder $\PP$
and any countable family $\{\mathcal{D}_n : n\in\mathbb{N}\}$ of dense
subsets of $\PP$ there is a filter intersecting every
$\mathcal{D}_i$. Thus, there are generic filters $G$ for countable
transitive models. In general, no such $G$ belongs to $M$ and
therefore the extension $M[G]$ is proper. We formalize the proof of
RSL in Section~\ref{sec:rasiowa-sikorski-lemma}. A requisite result on
a version of the Axiom of Choice is formalized in
Section~\ref{sec:sequence-version-dc}. We then apply RSL to prove the
existence of generic filters in Section~\ref{sec:generic-filter}.

Every  $y \in M[G]$ is obtained from an element $\dot y$ of $M$, thus
elements of $M$ are construed as \emph{names} or codes for elements of
$M[G]$.
The decoding is given by the function
$\val$, which takes the generic filter $G$ as a parameter. To
prove that $M$ is contained in $M[G]$ it suffices to give a name for
each element of $M$; we define the function $\chk$ which assigns
a name for each $x\in M$. Showing that $\chk(x)\in M$
when $x\in M$ involves some technical issues that will
be addressed in a further work. We explain names, $\val$, and
$\chk$ in Section~\ref{sec:names}.

A central part of this formalization project involves showing that
$\ZFC$ holds in the generic extension. This is most relevant since
forcing is essentially the only known way to \emph{extend} models of
$\ZFC$ (while preserving ordinals). The most difficult step to achieve
this goal is to define the \emph{forcing relation}, that allows to
connect satisfaction in $M$ to that of $M[G]$; this is needed to show
that the Powerset axiom and the axiom schemes of Separation and
Replacement hold in $M[G]$. In
Section~\ref{sec:pairing-generic-extension}
 we tackle the Pairing
Axiom. This does not require the forcing relation, but provides an
illustration of the use of names. The development can be downloaded
from \url{https://cs.famaf.unc.edu.ar/~mpagano/forcing/} and it can
also be found among the source files of this arXiv version.

\subsection{Related work}\label{sec:related-work}
Formalization of mathematics serves many purposes
\cite{simpson-theorem-proving-math}. The most obvious one is to
increase reliability in a result and/or its proof. This has been the
original motivation that lead Voevodsky to gather many researchers
around \emph{homotopy type theory} and its formalization in Coq
\cite{hottbook}; the same applies to the four color theorem (checked
by Gonthier \cite{MR2463991}) and the formidable \emph{Flyspeck}
project \cite{MR3659768} by the team conducted by Hales.

In our particular case, forcing and the set theoretic techniques that
are being formalized can be regarded as a mature technology and thus
the main goal is not to increase confidence. Nevertheless, the level
of detail in a formalization of this sort always provides additional
information about the inner workings of the theory: It is expected, for
instance, to have a detailed account of which axioms are necessary to
define and use forcing. Finally, we support the vision that a growing
corpus of formalized mathematics can be a useful library for the
future generations. The question of how to systematize this corpus is
an ongoing project by Paulson \cite{ALEXANDRIA}.

We will now discuss very succinctly recent formalizations of
set theory and forcing. The closest formalizations are those based on
Isabelle. Let us remark that Isabelle allows for different logical
foundations; in particular, Paulson carried out his formalizations
on top of Isabelle/FOL which is based on first-order logic.

There is another major framework in Isabelle based on higher order
logic, Isabelle/HOL. This framework is very active, and as a 
consequence more automated tools are available. Isabelle/HOL has 
basic chapters on set theory. One of those, by Steven Obua, proceeds up to
well founded relations and provides translations between types in HOL
(for instance \isatt{nat}) to  sets  (elements of type
\isatt{ZF}). Another one, by A.~Popescu and D.~Traytel, reaches
cardinal arithmetic. This is fairly limited for our purposes.

Concerning automation, B.~Zhan has developed a new tool called
\texttt{auto2} and applied it to untyped set theory
\cite{10.1007/978-3-319-66107-0_32}. He has redeveloped essentially
the basic results in Isabelle/ZF, but goes in a different
direction. Nevertheless, a majority of results in
Isabelle/ZF are not yet implemented using this tool, and  another
downside is that proofs using it do not follow the standard Isar
language (see Section~\ref{sec:isabellezf}).

As far as we know, there is little progress on formalizations of
forcing in type theory. Most relevant is the work by K.~Quirin
\cite{Quirin}, where a sheaf-theoretic initial approach to forcing is
implemented in Coq. This language is extremely different to the
standard approach of constructing models of $\ZFC$, and it might be
difficult (once the forcing machinery is set) to translate results in
the literature using ctms to this one. In any case, the translation to
set theory of what Quirin accomplishes is to define a generic
extension (where $\CH$ should fail) and to construct a set $K$ (a
candidate counterexample) and injections $\N\hookrightarrow K$ and
$K\hookrightarrow\R$. But the most important part, that is, that there
are no surjections $\N\twoheadrightarrow K$ and
$K\twoheadrightarrow\R$, is left for a future
work. 


\section{Isabelle/ZF}\label{sec:isabellezf}

Let us introduce briefly Paulson's formalization of ZF
\cite{paulson2017isabelle} in Isabelle and the main aspects of his
formal proof for the relative consistency of the Axiom of Choice
\cite{paulson_2003}; we will only focus on those aspects that are
essential to keep this paper self-contained, and refer the interested
reader to Paulson's articles.
Isabelle/ZF includes a development of classical first-order logic,
FOL. Both of them are  built upon the core library \emph{Pure}. 

In Isabelle/ZF sets are \emph{individuals}, i.e.\ terms of type
\isatt{i} and formulas have type \isatt{o} (akin to a \emph{Bool}
type, but at the object level).  The axiomatization of $\ZFC$ in
Isabelle/ZF proceeds by postulating a binary predicate
\isatt{\ensuremath{\in}} and several set constructors (terms and
functions with values in \isatt{i}) corresponding to the empty set (the
constant \isatt{\isadigit{0}}), powersets, and one further constant
\isatt{inf} for an infinite set. The axioms, being formulas, are terms
of type \isatt{o}; the foundation axiom, for example, is formalized as
(the universal closure of) \isa{{\isachardoublequoteopen}A\
  {\isacharequal}\ {\isadigit{0}}\ {\isasymor}%
  \ {\isacharparenleft}{\isasymexists}x{\isasymin}A{\isachardot}\ %
  {\isasymforall}y{\isasymin}x{\isachardot}\ %
  y{\isasymnotin}A{\isacharparenright}%
  {\isachardoublequoteclose}}. %
Besides the axioms, Isabelle/ZF also introduces several definitions
(for example, pairs and sets defined by comprehension using
separation) and syntactic abbreviations to keep the formalization
close to the customary manner of doing mathematics.  Working with the
library and extending it is quite straightforward.  As an example, we
introduce a new term-former (which is a combination of instances of
replacement and separation) denoting the image of a function over a
set defined by comprehension, namely
$\{b(x): x\in A\text{ and }Q(x)\}$:
\begin{isabelle}
  \isacommand{definition}\isamarkupfalse%
  \ SepReplace\ {\isacharcolon}{\isacharcolon}\
  {\isachardoublequoteopen}{\isacharbrackleft}i{\isacharcomma}\
  i{\isasymRightarrow}i{\isacharcomma}\ i{\isasymRightarrow}\
  o{\isacharbrackright}\
  {\isasymRightarrow}i{\isachardoublequoteclose}\
  \isakeyword{where}\isanewline \ \
  {\isachardoublequoteopen}SepReplace{\isacharparenleft}A{\isacharcomma}b{\isacharcomma}Q{\isacharparenright}\
  {\isacharequal}{\isacharequal}\ {\isacharbraceleft}y\ {\isachardot}\
  x{\isasymin}A{\isacharcomma}\
  y{\isacharequal}b{\isacharparenleft}x{\isacharparenright}\
  {\isasymand}\
  Q{\isacharparenleft}x{\isacharparenright}{\isacharbraceright}{\isachardoublequoteclose}
\end{isabelle}
\noindent %
We are then able to add the abbreviation \isa{{\isacharbraceleft}b\
  {\isachardot}{\isachardot}\ x{\isasymin}A{\isacharcomma}\
  Q{\isacharbraceright}} as a notation for
\isa{SepReplace{\isacharparenleft}A{\isacharcomma}b{\isacharcomma}Q{\isacharparenright}}. The
characterization of our new constructor is given by
\begin{isabelle}
\isacommand{lemma}\isamarkupfalse%
\ Sep{\isacharunderscore}and{\isacharunderscore}Replace{\isacharcolon}\ {\isachardoublequoteopen}{\isacharbraceleft}b{\isacharparenleft}x{\isacharparenright}\ {\isachardot}{\isachardot}\ x{\isasymin}A{\isacharcomma}\ Q{\isacharparenleft}x{\isacharparenright}\ {\isacharbraceright}\ {\isacharequal}\ {\isacharbraceleft}b{\isacharparenleft}x{\isacharparenright}\ {\isachardot}\ x{\isasymin}{\isacharbraceleft}y{\isasymin}A{\isachardot}\ Q{\isacharparenleft}y{\isacharparenright}{\isacharbraceright}{\isacharbraceright}{\isachardoublequoteclose}
\end{isabelle}

We now discuss relativization in Isabelle/ZF. Relativized versions of the
 axioms can be found in the formalization of constructibility \cite{paulson_2003}. For
 example, the relativized Axiom of Foundation is
\begin{isabelle}
\isacommand{definition}\isamarkupfalse%
\ foundation{\isacharunderscore}ax\ {\isacharcolon}{\isacharcolon}\ {\isachardoublequoteopen}{\isacharparenleft}i{\isacharequal}{\isachargreater}o{\isacharparenright}\ {\isacharequal}{\isachargreater}\ o{\isachardoublequoteclose}\ \isakeyword{where}\isanewline
 {\isachardoublequoteopen}foundation{\isacharunderscore}ax{\isacharparenleft}M{\isacharparenright}\ {\isacharequal}{\isacharequal}\isanewline
  {\isasymforall}x{\isacharbrackleft}M{\isacharbrackright}{\isachardot}\ {\isacharparenleft}{\isasymexists}y{\isacharbrackleft}M{\isacharbrackright}{\isachardot}\ y{\isasymin}x{\isacharparenright}\ {\isasymlongrightarrow}\ {\isacharparenleft}{\isasymexists}y{\isacharbrackleft}M{\isacharbrackright}{\isachardot}\ y{\isasymin}x\ {\isacharampersand}\ {\isachartilde}{\isacharparenleft}{\isasymexists}z{\isacharbrackleft}M{\isacharbrackright}{\isachardot}\ z{\isasymin}x\ {\isacharampersand}\ z\ {\isasymin}\ y{\isacharparenright}{\isacharparenright}{\isachardoublequoteclose}
\end{isabelle}

\noindent The relativized quantifier
\isa{{\isasymforall}x{\isacharbrackleft}M{\isacharbrackright}{\isachardot}\
  P(x)} is a shorthand for \isa{{\isasymforall}x{\isachardot}\ M(x)
  {\isasymlongrightarrow} P(x)}. In order to express that a (set) model
satisfies this axiom we use  the ``coercion''
\isa{{\isacharhash}{\isacharhash} :: i ={\isachargreater} (i
  ={\isachargreater} o)} (that maps a set $A$ to the predicate
$\lambda x . (x\in A)$) provided by Isabelle/ZF. As a trivial example we
can show that the empty set satisfies Foundation:
\begin{isabelle}
\isacommand{lemma}\isamarkupfalse%
\ emp{\isacharunderscore}foundation\ {\isacharcolon}\ {\isachardoublequoteopen}foundation{\isacharunderscore}ax{\isacharparenleft}{\isacharhash}{\isacharhash}{\isadigit{0}}{\isacharparenright}{\isachardoublequoteclose}
\end{isabelle}

Mathematical texts usually start by fixing a context that defines
parameters and assumptions needed to develop theorems
and results. In Isabelle the way of defining contexts is through
\emph{locales}~\cite{ballarin2010tutorial}.
Locales can be combined and extended by adding more parameters and assuming
more facts, leading to a new locale. For example a context describing
lattices can be extended to distributive lattices.
The way to instantiate a locale is by \emph{interpreting} it, which consists
of giving concrete values to parameters and proving the assumptions.
In our work, we use locales to organize the formalization and to make
explicit the assumptions of the most important results.

Let us close this section with a brief comment about the facilities
provided by the Isabelle framework. The edition is done in an IDE
called \texttt{jEdit}, which is bundled with the standard Isabelle
distribution; it offers the user a fair amount of tools in order to
manage theory files, searching for theorems and concepts spread
through the source files, and includes tracing utilities for the
automatic tools. A main feature is a window showing the \emph{proof
  state}, where the active (sub)goals are shown, along with the
already obtained results and possibly errors.

Isabelle proofs can be written in two dialects.  The older one, and
also more basic, follows a procedural approach, where one applies
several tactics in order to decompose the goal into simpler ones and
then solving them (with the aid of automation); the original work by
Paulson used this method. Under this approach proofs are constructed
top-down resulting in proof-scripts that conceal the mathematical
reasoning behind the proof, since the intermediate steps are only
shown in the proof state. For this reason, the proof language
\emph{Isar} was developed, starting with Wenzel's
work~\cite{DBLP:conf/tphol/Wenzel99}. Isar is mostly declarative, and
its main purpose is to construct \emph{proof documents} that (in
principle) can be read and understood without the need of running the
code.

We started this development using the procedural approach, but soon
after we realized that for our purposes the Isar language was far more
appropriate.


\section{Forcing notions}\label{sec:forcing-notions}

In this section we present a proof of the Rasiowa-Sikorski lemma which
uses the principle of dependent choices. We start by introducing
the necessary definitions about preorders; then, we explain and prove
the principle of dependent choice most suitable for our purpose.

It is to be noted that the order of presentation of the material
deviates a bit from the dependency of the source  files. The
file containing the most basic results and definitions that follow
imports that containing the results of
Subsection~\ref{sec:sequence-version-dc}.

\begin{definition}
  A preorder on a set $P$ is a binary relation ${\leqslant}$ which is
  reflexive and transitive.
\end{definition}

The preorder relation will be represented as a set of pairs, and hence
it is a term of type
\isatt{i}.
\begin{definition}
  Given a preorder $(P,\leqslant)$ we say that two elements $p,q$ are
  \emph{compatible} if they have a lower bound in $P$. Notice that
  the elements of $P$ are also sets, therefore they have type
  \isatt{i}.
  \begin{isabelle}%
  \isacommand{definition}\isamarkupfalse%
\ compat{\isacharunderscore}in\ {\isacharcolon}{\isacharcolon}\ {\isachardoublequoteopen}i{\isasymRightarrow}i{\isasymRightarrow}i{\isasymRightarrow}i{\isasymRightarrow}o{\isachardoublequoteclose}\ \isakeyword{where}\isanewline
\ \ {\isachardoublequoteopen}compat{\isacharunderscore}in{\isacharparenleft}P{\isacharcomma}leq{\isacharcomma}p{\isacharcomma}q{\isacharparenright}\ {\isacharequal}{\isacharequal}\ {\isasymexists}d{\isasymin}P\ {\isachardot}\ {\isasymlangle}d{\isacharcomma}p{\isasymrangle}{\isasymin}leq\ {\isasymand}\ {\isasymlangle}d{\isacharcomma}q{\isasymrangle}{\isasymin}leq{\isachardoublequoteclose}
\end{isabelle}
\end{definition}

\begin{definition}
  A \emph{forcing notion} is a preorder $(P,\leqslant)$ with a maximal element $\mathbbm{1} \in P$.
  \begin{isabelle}
\isacommand{locale}\isamarkupfalse%
\ forcing{\isacharunderscore}notion\ {\isacharequal}\isanewline
\ \ \isakeyword{fixes}\ P\ leq\ one\isanewline
\ \ \isakeyword{assumes}\ one{\isacharunderscore}in{\isacharunderscore}P{\isacharcolon}  \ {\isachardoublequoteopen}one\ {\isasymin}\ P{\isachardoublequoteclose}\isanewline
 \ \ \isakeyword{and}\ leq{\isacharunderscore}preord{\isacharcolon} \ \ \ {\isachardoublequoteopen}preorder{\isacharunderscore}on{\isacharparenleft}P{\isacharcomma}leq{\isacharparenright}{\isachardoublequoteclose}\isanewline
 \ \ \isakeyword{and}\ one{\isacharunderscore}max{\isacharcolon}  \ \ {\isachardoublequoteopen}{\isasymforall}p{\isasymin}P{\isachardot}\ {\isasymlangle}p{\isacharcomma}one{\isasymrangle}{\isasymin}leq{\isachardoublequoteclose}
\end{isabelle}
\end{definition}
\noindent The locale \isatt{forcing{\isacharunderscore}notion}  introduces a mathematical
context where we work assuming the forcing notion
$(P,\leqslant, \mathbbm{1})$. 
In the following definitions we are in
the locale \isatt{forcing{\isacharunderscore}notion}.

A set $D$ is \emph{dense} if every element $p\in P$ has a lower bound
in $D$ and there is also a weaker definition which asks for a lower
bound in $D$ only for the elements below some fixed element $q$. 
\begin{isabelle}
  \isacommand{definition}\isamarkupfalse%
\ dense\ {\isacharcolon}{\isacharcolon}\ {\isachardoublequoteopen}i{\isasymRightarrow}o{\isachardoublequoteclose}\ \isakeyword{where}\isanewline
\ \ {\isachardoublequoteopen}dense{\isacharparenleft}D{\isacharparenright}\ {\isacharequal}{\isacharequal}\ {\isasymforall}p{\isasymin}P{\isachardot}\ {\isasymexists}d{\isasymin}D\ {\isachardot}\ {\isasymlangle}d{\isacharcomma}p{\isasymrangle}{\isasymin}leq{\isachardoublequoteclose}\isanewline
\isanewline
\isacommand{definition}\isamarkupfalse%
\ dense{\isacharunderscore}below\ {\isacharcolon}{\isacharcolon}\ {\isachardoublequoteopen}i{\isasymRightarrow}i{\isasymRightarrow}o{\isachardoublequoteclose}\ \isakeyword{where}\isanewline
\ \ {\isachardoublequoteopen}dense{\isacharunderscore}below{\isacharparenleft}D{\isacharcomma}q{\isacharparenright}\ {\isacharequal}{\isacharequal}\ {\isasymforall}p{\isasymin}P{\isachardot}\ {\isasymlangle}p{\isacharcomma}q{\isasymrangle}{\isasymin}leq\ {\isasymlongrightarrow}\ {\isacharparenleft}{\isasymexists}d{\isasymin}D\ {\isachardot}\ {\isasymlangle}d{\isacharcomma}p{\isasymrangle}{\isasymin}leq{\isacharparenright}{\isachardoublequoteclose}
\end{isabelle}
Since
the relation $\leqslant$ is reflexive, it is obvious that $P$ is
dense. Actually, this follows automatically once the appropriate definitions are
unfolded:
\begin{isabelle}
\isacommand{lemma}\isamarkupfalse%
\ P{\isacharunderscore}dense{\isacharcolon}\ {\isachardoublequoteopen}dense{\isacharparenleft}P{\isacharparenright}{\isachardoublequoteclose}\isanewline
%
\ \ %
%
\isacommand{using}\isamarkupfalse%
\ leq{\isacharunderscore}preord\isanewline
\ \ \isacommand{unfolding}\isamarkupfalse%
\ preorder{\isacharunderscore}on{\isacharunderscore}def\ refl{\isacharunderscore}def\ dense{\isacharunderscore}def\isanewline
\ \ \isacommand{by}\isamarkupfalse%
\ blast%
\end{isabelle}
Here, the automatic tactic \isa{blast} solves the goal. In the
procedural approach, goals are refined with the command
\textbf{apply}~\emph{tactic}, and proofs are finished using \textbf{done}. 
Then \textbf{by $\dots$} is an idiom for 
\textbf{apply $\dots$ done}.
 
We say that $F\subseteq P$ is increasing (or upward closed) if every
extension of any element in $F$ is also in $F$.
\begin{isabelle}
\isacommand{definition}\isamarkupfalse%
\ increasing\ {\isacharcolon}{\isacharcolon}\ {\isachardoublequoteopen}i{\isasymRightarrow}o{\isachardoublequoteclose}\ \isakeyword{where}\isanewline
\ \ {\isachardoublequoteopen}increasing{\isacharparenleft}F{\isacharparenright}\ {\isacharequal}{\isacharequal}\ {\isasymforall}x{\isasymin}F{\isachardot}\ {\isasymforall}\ p{\isasymin}P\ {\isachardot}\ {\isasymlangle}x{\isacharcomma}p{\isasymrangle}{\isasymin}leq\ {\isasymlongrightarrow}\ p{\isasymin}F{\isachardoublequoteclose}
\end{isabelle}
A filter is an increasing set $G$ with all its elements being compatible in $G$.
\begin{isabelle}
\isacommand{definition}\isamarkupfalse%
\ filter\ {\isacharcolon}{\isacharcolon}\ {\isachardoublequoteopen}i{\isasymRightarrow}o{\isachardoublequoteclose}\ \isakeyword{where}\isanewline
\ \ {\isachardoublequoteopen}filter{\isacharparenleft}G{\isacharparenright}\ {\isacharequal}{\isacharequal}\ G{\isasymsubseteq}P\ {\isasymand}\ increasing{\isacharparenleft}G{\isacharparenright}\ {\isasymand}\isanewline \ \ {\isacharparenleft}{\isasymforall}p{\isasymin}G{\isachardot}\ {\isasymforall}q{\isasymin}G{\isachardot}\ compat{\isacharunderscore}in{\isacharparenleft}G{\isacharcomma}leq{\isacharcomma}p{\isacharcomma}q{\isacharparenright}{\isacharparenright}{\isachardoublequoteclose}
\end{isabelle}

We finally introduce the upward closure of a set
and prove that the closure of $A$ is a filter if its elements are
compatible in $A$.
\begin{isabelle}
\isacommand{definition}\isamarkupfalse%
\ upclosure\ {\isacharcolon}{\isacharcolon}\ {\isachardoublequoteopen}i{\isasymRightarrow}i{\isachardoublequoteclose}\ \isakeyword{where}\isanewline
\ \ {\isachardoublequoteopen}upclosure{\isacharparenleft}A{\isacharparenright}\ {\isacharequal}{\isacharequal}\ {\isacharbraceleft}p{\isasymin}P{\isachardot}{\isasymexists}a{\isasymin}A{\isachardot}{\isasymlangle}a{\isacharcomma}p{\isasymrangle}{\isasymin}leq{\isacharbraceright}{\isachardoublequoteclose}\isanewline
\isacommand{lemma}\isamarkupfalse%
\ \ closure{\isacharunderscore}compat{\isacharunderscore}filter{\isacharcolon}
\ \ {\isachardoublequoteopen}A{\isasymsubseteq}P\ {\isasymLongrightarrow}\isanewline\ \ {\isacharparenleft}{\isasymforall}p{\isasymin}A{\isachardot}{\isasymforall}q{\isasymin}A{\isachardot}\ compat{\isacharunderscore}in{\isacharparenleft}A{\isacharcomma}leq{\isacharcomma}p{\isacharcomma}q{\isacharparenright}{\isacharparenright}\ {\isasymLongrightarrow}\ filter{\isacharparenleft}upclosure{\isacharparenleft}A{\isacharparenright}{\isacharparenright}{\isachardoublequoteclose}
\end{isabelle}
As usual
with procedural proofs, the refinement process goes ``backwards,''
from the main goal to simpler ones. The proof of this last lemma takes
21 lines and 34 proof commands and is one of the longest procedural
proofs in the development.  It was  at
the moment of its implementation that we realized that a declarative
approach was best because, apart from being more readable, the
reasoning flows mostly in a forward fashion.


\subsection{A sequence version of Dependent
  Choices}\label{sec:sequence-version-dc} 
The Rasiowa-Sikorski lemma follows naturally from a ``pointed''
version of   the \emph{Principle of Dependent Choices ($\DC$)} which,
in turn, is a consequence of the Axiom of Choice ($\AC$). It is
therefore natural to take as a starting point the theory \isatt{AC}
which adds the latter axiom to the toolkit of Isabelle/ZF.

The statement we are interested in is the following:
\begin{quote}
  (Pointed $\DC$) Let $\rr$ be a binary relation on $A$, and $a\in A$. If
  $\forall x\in A.\,  \exists y\in A.\, x\rr y$, then there exists
  $f:\om\to A$ such that $f(0)=a$ and $f(n)\rr f(n+1)$ for all
  $n\in\om$.
\end{quote}

Two different versions of $\DC$ (called $\DC0$ and $\DC(\kappa)$) have
already been formalized by Krzysztof
Grabczewski~\cite{DBLP:journals/jar/PaulsonG96}, as part of a study of
equivalents of $\AC$ (following Rubin and Rubin
\cite{rubin1985equivalents}). Nevertheless, those are not convenient
for our purposes. In fact, the axiom $\DC0$ corresponds essentially to
our Pointed $\DC$ but without the constraint $f(0)=a$; it is a nice
exercise to show that $\DC0$ implies Pointed $\DC$, but a
formalization would have a moderate length. On the other hand,
$\DC(\kappa)$ is rather different in nature and it is tailored to
obtain another proposition equivalent to the axiom of choice
(actually,
$\AC\longleftrightarrow (\forall \kappa .\;
\mathrm{Card}(\kappa)\longrightarrow \DC(\kappa))$).  Finally, the
shortest path from $\AC$ 
to $\DC0$ using already formalized
material involves a complicated detour (130+ proof commands spanning
various files of the \isatt{ZF-AC} theory and going through the Well
Ordering Theorem and  $\DC(\omega)$), compared to the mere 11 commands from $\AC$ to
\isatt{AC{\isacharunderscore}func{\isacharunderscore}Pow}. This last
one is the
choice principle that we use in our formalization of Pointed $\DC$, and states 
the existence of
choice functions (``selectors'') on $\P(A)\sm \{\mty\}$):
\[
\exists (s: \P(A)\sm \{\mty\}\to A). \, \forall X\sbq A.\ X\neq \mty \limp
s(X) \in X.
\]
Another advantage of taking
\isatt{AC{\isacharunderscore}func{\isacharunderscore}Pow} as a
starting point is that it does not involve proper classes: The version
of  $\AC$ in Isabelle/ZF corresponds to an axiom scheme of first-order
logic and as such is not a standard formulation. 

The strategy to prove Pointed $\DC$ (following a proof in Moschovakis
\cite{moschovakis1994notes}) is to define the function $f$ discussed above by
primitive recursion on the naturals, which can be done easily thanks
to the package of Isabelle/ZF
\cite{paulson1995set,paulson2000fixedpoint} for definitions by
recursion on inductively defined sets.\footnote{The package figures
  out the inductive set at hand and checks that the recursive
  definition makes sense; for example, it rejects definitions with a
  missing case.}

\begin{isabelle}\isamarkuptrue%
\isacommand{consts}\isamarkupfalse%
\ dc{\isacharunderscore}witness\ {\isacharcolon}{\isacharcolon}\ {\isachardoublequoteopen}i\ {\isasymRightarrow}\ i\ {\isasymRightarrow}\ i\ {\isasymRightarrow}\ i\ {\isasymRightarrow}\ i\ {\isasymRightarrow}\ i{\isachardoublequoteclose}\isanewline
\isacommand{primrec}\isamarkupfalse%
\isanewline
\ \ wit{\isadigit{0}}\ \ \ {\isacharcolon}\ {\isachardoublequoteopen}dc{\isacharunderscore}witness{\isacharparenleft}{\isadigit{0}}{\isacharcomma}A{\isacharcomma}a{\isacharcomma}s{\isacharcomma}R{\isacharparenright}\ {\isacharequal}\ a{\isachardoublequoteclose}\isanewline
\ \ witrec\ {\isacharcolon}\ {\isachardoublequoteopen}dc{\isacharunderscore}witness{\isacharparenleft}succ{\isacharparenleft}n{\isacharparenright}{\isacharcomma}A{\isacharcomma}a{\isacharcomma}s{\isacharcomma}R{\isacharparenright}\ {\isacharequal}\isanewline \ \ \ \ \ \ \ \ \ \  \ \ s{\isacharbackquote}{\isacharbraceleft}x{\isasymin}A{\isachardot}\ {\isasymlangle}dc{\isacharunderscore}witness{\isacharparenleft}n{\isacharcomma}A{\isacharcomma}a{\isacharcomma}s{\isacharcomma}R{\isacharparenright}{\isacharcomma}x{\isasymrangle}{\isasymin}R\ {\isacharbraceright}{\isachardoublequoteclose}
\end{isabelle}

Besides the natural argument and the parameters $A$, $a$, and $R$, the
function \isatt{dc{\isacharunderscore}witness} has a function $s$ as a
parameter. If this function is a selector for $\P(A)\sm \{\mty\}$, the
function
$f(n)\defi {}$\isatt{dc{\isacharunderscore}witness}$(n,A,a,s,R)$ will
satify $\DC$. Notice that $s$ is a term of type \isatt{i} (a function
construed as a set of pairs) and an expression
\isatt{s{\isacharbackquote}b} is notation for \isatt{apply(s,b)},
where \isatt{apply\ {\isacharcolon}{\isacharcolon}\
  {\isachardoublequoteopen}i\ {\isasymRightarrow}\ i\
  {\isasymRightarrow}\ i{\isachardoublequoteclose}} is the operation
of function application.

The proof is mostly routine; after a few lemmas (26 proof
commands in total) we obtain the
following theorem:

\begin{isabelle}
\isacommand{theorem}\isamarkupfalse%
\ pointed{\isacharunderscore}DC\ \ {\isacharcolon}\ {\isachardoublequoteopen}{\isacharparenleft}{\isasymforall}x{\isasymin}A{\isachardot}\ {\isasymexists}y{\isasymin}A{\isachardot}\ {\isasymlangle}x{\isacharcomma}y{\isasymrangle}{\isasymin}\ R{\isacharparenright}\ {\isasymLongrightarrow}\isanewline
     \ \ \ {\isasymforall}a{\isasymin}A{\isachardot}\ {\isacharparenleft}{\isasymexists}f\ {\isasymin}\ nat{\isasymrightarrow}A{\isachardot}\ f{\isacharbackquote}{\isadigit{0}}\ {\isacharequal}\ a\ {\isasymand}\ {\isacharparenleft}{\isasymforall}n\ {\isasymin}\ nat{\isachardot}\ {\isasymlangle}f{\isacharbackquote}n{\isacharcomma}f{\isacharbackquote}succ{\isacharparenleft}n{\isacharparenright}{\isasymrangle}{\isasymin}R{\isacharparenright}{\isacharparenright}{\isachardoublequoteclose}
\end{isabelle}

We need a further, ``diagonal'' version of $\DC$  to prove
Rasiowa-Sikorski. That is, if the assumption holds for a sequence of
relations $S_n$,  then $f(n) \mathrel{S_{n+1}} f(n+1)$ for all $n$.

We first obtain a corollary of $\DC$ changing $A$ for
$A\times\mathtt{nat}$, whose procedural proof takes 16 lines:

\begin{isabelle}
\isacommand{corollary}\isamarkupfalse%
\ DC{\isacharunderscore}on{\isacharunderscore}A{\isacharunderscore}x{\isacharunderscore}nat\ {\isacharcolon}\ \isanewline
\ \ {\isachardoublequoteopen}{\isacharparenleft}{\isasymforall}x{\isasymin}A{\isasymtimes}nat{\isachardot}\ {\isasymexists}y{\isasymin}A{\isachardot}\ {\isasymlangle}x{\isacharcomma}{\isasymlangle}y{\isacharcomma}succ{\isacharparenleft}snd{\isacharparenleft}x{\isacharparenright}{\isacharparenright}{\isasymrangle}{\isasymrangle}\ {\isasymin}\ R{\isacharparenright}\ {\isasymLongrightarrow}\isanewline
\ \ \ \ {\isasymforall}a{\isasymin}A{\isachardot}\ {\isacharparenleft}{\isasymexists}f\ {\isasymin}\ nat{\isasymrightarrow}A{\isachardot}\ f{\isacharbackquote}{\isadigit{0}}\ {\isacharequal}\ a\ {\isasymand}\isanewline
\ \ \ \ \ \ \ \ \ \ \ \ {\isacharparenleft}{\isasymforall}n\ {\isasymin}\ nat{\isachardot}\ {\isasymlangle}{\isasymlangle}f{\isacharbackquote}n{\isacharcomma}n{\isasymrangle}{\isacharcomma}{\isasymlangle}f{\isacharbackquote}succ{\isacharparenleft}n{\isacharparenright}{\isacharcomma}succ{\isacharparenleft}n{\isacharparenright}{\isasymrangle}{\isasymrangle}{\isasymin}R{\isacharparenright}{\isacharparenright}{\isachardoublequoteclose}
\end{isabelle}

The following lemma is then proved automatically:

\begin{isabelle}
\isacommand{lemma}\isamarkupfalse%
\ aux{\isacharunderscore}sequence{\isacharunderscore}DC\ {\isacharcolon}\ {\isachardoublequoteopen}{\isasymforall}x{\isasymin}A{\isachardot}\ {\isasymforall}n{\isasymin}nat{\isachardot}\ {\isasymexists}y{\isasymin}A{\isachardot}\ {\isasymlangle}x{\isacharcomma}y{\isasymrangle}\ {\isasymin}\ S{\isacharbackquote}n\ {\isasymLongrightarrow}\isanewline
 \ \ {\isasymforall}x{\isasymin}A{\isasymtimes}nat{\isachardot}\ {\isasymexists}y{\isasymin}A{\isachardot}\isanewline
 \ \ \ {\isasymlangle}x{\isacharcomma}{\isasymlangle}y{\isacharcomma}succ{\isacharparenleft}snd{\isacharparenleft}x{\isacharparenright}{\isacharparenright}{\isasymrangle}{\isasymrangle}\ {\isasymin}\ {\isacharbraceleft}{\isasymlangle}{\isasymlangle}w{\isacharcomma}n{\isasymrangle}{\isacharcomma}{\isasymlangle}y{\isacharcomma}m{\isasymrangle}{\isasymrangle}{\isasymin}{\isacharparenleft}A{\isasymtimes}nat{\isacharparenright}{\isasymtimes}{\isacharparenleft}A{\isasymtimes}nat{\isacharparenright}{\isachardot}\ {\isasymlangle}w{\isacharcomma}y{\isasymrangle}{\isasymin}S{\isacharbackquote}m\ {\isacharbraceright}{\isachardoublequoteclose}\isanewline
\ \ %
\isacommand{by}\isamarkupfalse%
\ auto%
\end{isabelle}
And after a short proof we arrive to $\DC$ for a sequence of relations:

\begin{isabelle}
\isacommand{lemma}\isamarkupfalse%
\ sequence{\isacharunderscore}DC{\isacharcolon}\ {\isachardoublequoteopen}{\isasymforall}x{\isasymin}A{\isachardot}\ {\isasymforall}n{\isasymin}nat{\isachardot}\ {\isasymexists}y{\isasymin}A{\isachardot}\ {\isasymlangle}x{\isacharcomma}y{\isasymrangle}\ {\isasymin}\ S{\isacharbackquote}n\ {\isasymLongrightarrow}\isanewline
\ \ \ \ {\isasymforall}a{\isasymin}A{\isachardot}\ {\isacharparenleft}{\isasymexists}f\ {\isasymin}\ nat{\isasymrightarrow}A{\isachardot}\ f{\isacharbackquote}{\isadigit{0}}\ {\isacharequal}\ a\ {\isasymand}\isanewline
\ \ \ \ \ \ \  {\isacharparenleft}{\isasymforall}n\ {\isasymin}\ nat{\isachardot}\ {\isasymlangle}f{\isacharbackquote}n{\isacharcomma}f{\isacharbackquote}succ{\isacharparenleft}n{\isacharparenright}{\isasymrangle}{\isasymin}S{\isacharbackquote}succ{\isacharparenleft}n{\isacharparenright}{\isacharparenright}{\isacharparenright}{\isachardoublequoteclose}\isanewline
%
\ \ %
%
\isacommand{apply}\isamarkupfalse%
\ {\isacharparenleft}drule\ aux{\isacharunderscore}sequence{\isacharunderscore}DC{\isacharparenright}\isanewline
\ \ \isacommand{apply}\isamarkupfalse%
\ {\isacharparenleft}drule\ DC{\isacharunderscore}on{\isacharunderscore}A{\isacharunderscore}x{\isacharunderscore}nat{\isacharcomma}\ auto{\isacharparenright}\isanewline
\ \ \isacommand{done}\isamarkupfalse%
\end{isabelle}

\subsection{The Rasiowa-Sikorski lemma}\label{sec:rasiowa-sikorski-lemma}
In order to state this Lemma, we gather the relevant hypotheses into a locale:

\begin{isabelle}%
\isacommand{locale}\isamarkupfalse%
\ countable{\isacharunderscore}generic\ {\isacharequal}\ forcing{\isacharunderscore}notion\ {\isacharplus}\isanewline
\ \ \isakeyword{fixes}\ {\isasymD}\isanewline
\ \ \isakeyword{assumes}\ countable{\isacharunderscore}subs{\isacharunderscore}of{\isacharunderscore}P{\isacharcolon}\ \ {\isachardoublequoteopen}{\isasymD}\ {\isasymin}\ nat{\isasymrightarrow}Pow{\isacharparenleft}P{\isacharparenright}{\isachardoublequoteclose}\isanewline
\ \ \isakeyword{and} \ seq{\isacharunderscore}of{\isacharunderscore}denses{\isacharcolon}  {\isachardoublequoteopen}{\isasymforall}n\ {\isasymin}\ nat{\isachardot}\ dense{\isacharparenleft}{\isasymD}{\isacharbackquote}n{\isacharparenright}{\isachardoublequoteclose}
\end{isabelle}
That is, $\calD$ is a sequence of dense subsets of the poset $P$. A
filter is \emph{$\calD$-generic} if it intersects every dense set in
the sequence.

\begin{isabelle}%
\isacommand{definition}\isamarkupfalse%
\ \ D{\isacharunderscore}generic\ {\isacharcolon}{\isacharcolon}\ {\isachardoublequoteopen}i{\isasymRightarrow}o{\isachardoublequoteclose}\ \isakeyword{where}\isanewline
\ \ {\isachardoublequoteopen}D{\isacharunderscore}generic{\isacharparenleft}G{\isacharparenright}\ {\isacharequal}{\isacharequal}\ filter{\isacharparenleft}G{\isacharparenright}\ {\isasymand}\ {\isacharparenleft}{\isasymforall}n{\isasymin}nat{\isachardot}{\isacharparenleft}{\isasymD}{\isacharbackquote}n{\isacharparenright}{\isasyminter}G{\isasymnoteq}{\isadigit{0}}{\isacharparenright}{\isachardoublequoteclose}
\end{isabelle}

We can now state the Rasiowa-Sikorski Lemma.
\begin{isabelle}%
\isacommand{theorem}\isamarkupfalse%
\ rasiowa{\isacharunderscore}sikorski{\isacharcolon}\isanewline
\ \ {\isachardoublequoteopen}p{\isasymin}P\ {\isasymLongrightarrow}\ {\isasymexists}G{\isachardot}\ p{\isasymin}G\ {\isasymand}\ D{\isacharunderscore}generic{\isacharparenleft}G{\isacharparenright}{\isachardoublequoteclose}
\end{isabelle}

The intuitive argument for the result is simple: Once $p_0=p\in P$ is
fixed, we can recursively choose $p_{n+1}$ such that 
$p_n \geq p_{n+1}\in \calD_n$, since $\calD_n$ is dense in $P$. Then
the filter generated by $\{p_n : n\in \om\}$ intersects each set in
the sequence $\{\calD_n\}_n$. This argument appeals to the sequence
version of $\DC$; we have to prove first that the relevant relation
satisfies its hypothesis:

\begin{isabelle}%
\isacommand{lemma}\isamarkupfalse%
\  RS{\isacharunderscore}relation{\isacharcolon}\isanewline
\ \ \isakeyword{assumes}\isanewline
  {\isadigit{1}}{\isacharcolon}\ \ {\isachardoublequoteopen}x{\isasymin}P{\isachardoublequoteclose}\isanewline
   \isakeyword{and}\isanewline
  {\isadigit{2}}{\isacharcolon}\ \ {\isachardoublequoteopen}n{\isasymin}nat{\isachardoublequoteclose}\isanewline
\ \ \isakeyword{shows}\isanewline
   {\isachardoublequoteopen}{\isasymexists}y{\isasymin}P{\isachardot}\ {\isasymlangle}x{\isacharcomma}y{\isasymrangle}\ {\isasymin}\ {\isacharparenleft}{\isasymlambda}m{\isasymin}nat{\isachardot}\ {\isacharbraceleft}{\isasymlangle}x{\isacharcomma}y{\isasymrangle}{\isasymin}P{\isacharasterisk}P{\isachardot}\ {\isasymlangle}y{\isacharcomma}x{\isasymrangle}{\isasymin}leq\ {\isasymand}\ y{\isasymin}{\isasymD}{\isacharbackquote}{\isacharparenleft}pred{\isacharparenleft}m{\isacharparenright}{\isacharparenright}{\isacharbraceright}{\isacharparenright}{\isacharbackquote}n{\isachardoublequoteclose}
\end{isabelle}
These two proofs have been implemented  using the
Isar proof language.


\section{The generic extension}

Cohen's technique of forcing consists of constructing new models of
$\ZFC$ by adding a \emph{generic} subset $G$ of the forcing notion $P$
(a preorder with top). Given a model $M$ of $\ZFC$, the extension with
the generic subset $G$ is called \emph{the generic extension} of $M$,
denoted $M[G]$.  In this section we introduce all the necessary
concepts and results for defining $M[G]$; namely, we show, using
Rasiowa-Sikorski, that every preorder in a ctm admits a generic filter
and also develop the machinery of names. As an application of the
latter, we prove some basic
results about the generic extension.

\subsection{The generic filter}
\label{sec:generic-filter}
The following locale gathers the data needed to ensure the 
existence of an $M$-generic filter for a poset \isa{P}. 

\begin{isabelle}
\isacommand{locale}\isamarkupfalse%
\ forcing{\isacharunderscore}data\ {\isacharequal}\ forcing{\isacharunderscore}notion\ {\isacharplus}\isanewline
\ \ \isakeyword{fixes}\ M\ enum\isanewline
\ \ \isakeyword{assumes}\ M{\isacharunderscore}countable{\isacharcolon} \ \ {\isachardoublequoteopen}enum{\isasymin}bij{\isacharparenleft}nat{\isacharcomma}M{\isacharparenright}{\isachardoublequoteclose}\isanewline
 \ \ \isakeyword{and}\ P{\isacharunderscore}in{\isacharunderscore}M{\isacharcolon}  \ \ \ {\isachardoublequoteopen}P\ {\isasymin}\ M{\isachardoublequoteclose}\isanewline
 \ \ \isakeyword{and}\ leq{\isacharunderscore}in{\isacharunderscore}M{\isacharcolon}  \ {\isachardoublequoteopen}leq\ {\isasymin}\ M{\isachardoublequoteclose}\isanewline
 \ \ \isakeyword{and}\ trans{\isacharunderscore}M{\isacharcolon}  \ \ {\isachardoublequoteopen}Transset{\isacharparenleft}M{\isacharparenright}{\isachardoublequoteclose}
\end{isabelle}

An immediate consequence of the Rasiowa-Sikorski Lemma is the
existence of an $M$-generic filter for a poset \isa{P}.

\begin{isabelle}

\isacommand{lemma}\isamarkupfalse%
\ generic{\isacharunderscore}filter{\isacharunderscore}existence{\isacharcolon}\ \isanewline
\ \ {\isachardoublequoteopen}p{\isasymin}P\ {\isasymLongrightarrow}\ {\isasymexists}G{\isachardot}\ p{\isasymin}G\ {\isasymand}\ M{\isacharunderscore}generic{\isacharparenleft}G{\isacharparenright}{\isachardoublequoteclose}
\end{isabelle}

\noindent By defining an appropriate countable sequence of dense subsets of \isa{P},
\begin{isabelle}

\ \ \isacommand{let}\isamarkupfalse%
\isanewline
   \ \ {\isacharquery}D{\isacharequal}{\isachardoublequoteopen}{\isasymlambda}n{\isasymin}nat{\isachardot}\ {\isacharparenleft}if\ {\isacharparenleft}enum{\isacharbackquote}n{\isasymsubseteq}P\ {\isasymand}\ dense{\isacharparenleft}enum{\isacharbackquote}n{\isacharparenright}{\isacharparenright}\ \ then\ enum{\isacharbackquote}n\ else\ P{\isacharparenright}{\isachardoublequoteclose}
\end{isabelle}
\noindent we can instantiate the locale \isatt{countable{\isacharunderscore}generic}

\begin{isabelle}

\ \ \isacommand{have}\isamarkupfalse%
\ \isanewline
  \ Eq{\isadigit{2}}{\isacharcolon}\ {\isachardoublequoteopen}{\isasymforall}n{\isasymin}nat{\isachardot}\ {\isacharquery}D{\isacharbackquote}n\ {\isasymin}\ Pow{\isacharparenleft}P{\isacharparenright}{\isachardoublequoteclose}\isanewline
 \isacommand{by}\isamarkupfalse%
\ auto\isanewline
\ \ \isacommand{then}\isamarkupfalse%
\ \isacommand{have}\isamarkupfalse%
\isanewline
  \ Eq{\isadigit{3}}{\isacharcolon}\ {\isachardoublequoteopen}{\isacharquery}D{\isacharcolon}nat{\isasymrightarrow}Pow{\isacharparenleft}P{\isacharparenright}{\isachardoublequoteclose}\isanewline
 \isacommand{by}\isamarkupfalse%
\ {\isacharparenleft}rule\ lam{\isacharunderscore}codomain{\isacharparenright}\isanewline
\ \ \isacommand{have}\isamarkupfalse%
\isanewline
  \ Eq{\isadigit{4}}{\isacharcolon}\ {\isachardoublequoteopen}{\isasymforall}n{\isasymin}nat{\isachardot}\ dense{\isacharparenleft}{\isacharquery}D{\isacharbackquote}n{\isacharparenright}{\isachardoublequoteclose}
\end{isabelle}
\dots
\begin{isabelle}

\ \ \isacommand{from}\isamarkupfalse%
\ Eq{\isadigit{3}}\ \isakeyword{and}\ Eq{\isadigit{4}}\ \isacommand{interpret}\isamarkupfalse%
\ \isanewline
  \ \ cg{\isacharcolon}\ countable{\isacharunderscore}generic\ P\ leq\ one\ {\isacharquery}D\ \isanewline
 \isacommand{by}\isamarkupfalse%
\ {\isacharparenleft}unfold{\isacharunderscore}locales{\isacharcomma}\ auto{\isacharparenright}
\end{isabelle}
and then a  $\calD$-generic filter given by Rasiowa-Sikorski will be $M$-generic by construction. 

\begin{isabelle}

\ \ \isacommand{from}\isamarkupfalse%
\ cg{\isachardot}rasiowa{\isacharunderscore}sikorski\ \isakeyword{and}\ Eq{\isadigit{1}}\ \isacommand{obtain}\isamarkupfalse%
\ G\ \isakeyword{where}\ \isanewline
  \ Eq{\isadigit{6}}{\isacharcolon}\ {\isachardoublequoteopen}p{\isasymin}G\ {\isasymand}\ filter{\isacharparenleft}G{\isacharparenright}\ {\isasymand}\ {\isacharparenleft}{\isasymforall}n{\isasymin}nat{\isachardot}{\isacharparenleft}{\isacharquery}D{\isacharbackquote}n{\isacharparenright}{\isasyminter}G{\isasymnoteq}{\isadigit{0}}{\isacharparenright}{\isachardoublequoteclose}\isanewline
 \isacommand{unfolding}\isamarkupfalse%
\ cg{\isachardot}D{\isacharunderscore}generic{\isacharunderscore}def\ \isacommand{by}\isamarkupfalse%
\ blast\isanewline
\ \ \isacommand{then}\isamarkupfalse%
\ \isacommand{have}\isamarkupfalse%
\isanewline
  \ Eq{\isadigit{7}}{\isacharcolon}\ {\isachardoublequoteopen}{\isacharparenleft}{\isasymforall}D{\isasymin}M{\isachardot}\ D{\isasymsubseteq}P\ {\isasymand}\ dense{\isacharparenleft}D{\isacharparenright}{\isasymlongrightarrow}D{\isasyminter}G{\isasymnoteq}{\isadigit{0}}{\isacharparenright}{\isachardoublequoteclose}
\end{isabelle}

\noindent We omit the rest of this Isar proof.


\subsection{Names}
\label{sec:names}
We formalize the function $\val$ that allows to
construct the elements of the generic extension $M[G]$ from elements
of the ctm $M$ and the generic filter $G$. The definition of $\val$
can be written succinctly as a recursive equation
\begin{equation}\label{eq:def-val}
\val(G,\tau)\defi \{\val(G,\sigma) : \exists p\in \PP.\,
(\lb\sigma,p\rb\in\tau \y p\in G)\}.
\end{equation}
The justification that $\val$ is well-defined comes from a general
result (transfinite recursion on well-founded
relations~\cite[p. 48]{kunen2011set}). Given a well-founded relation
$R \subseteq A \times A$ and a functional
$H : A \times (A \to A) \to A$, the principle asserts the existence of
a function $F : A \to A$ satisfying
$F(a) = H(a,F\uparrow\,(R^{-1}(a)))$. This principle is formalized in
Isabelle/ZF and one can use the operator \isa{wfrec}\footnote{Notice
  that this form of recursive definitions is more general than the one
  used in the previous section to define
  \isa{dc{\isacharunderscore}witness}.} to define functions using
transfinite recursion. To be precise, \isa{wfrec :: [i, i, [i,i]=>i]
  => i} is a slight variation, where the first argument is the
relation, the third is the functional, and the second corresponds to
the argument of $F$. Notice that the relation and the function argument
of the functional are internalized as terms of type \isa{i}.

In our case the functional is called $Hv$ and takes an additional argument for the
parameter $G$:
\[
  \mathit{Hv}(G,y,f) = \{f(x) : x \in dom(y) \y \exists p\in \PP.\,
(\lb x,p\rb\in y \y p\in G)\}
\]
while the relation is given by:
\[
x \mathrel{\mathit{ed}} y \iff \exists p . \lb x,p\rb\in y.
\]
Recall that in $\ZFC$, an ordered pair $\lb x,y \rb$ is the set
$\{\{x\},\{x,y\}\}$. It is trivial to deduce the well-foundedness of
$\mathit{ed}$ from the fact that $\in$ is well-founded, which follows
from the Foundation Axiom.

In our formalization of this recursion,  the first argument of
\isa{wfrec} is the term of type  \isa{i} obtained by restricting the
relation $\mathit{ed}$ to  a set:
\begin{isabelle}
\isacommand{definition}\isamarkupfalse%
\isanewline
\ \ edrel\ {\isacharcolon}{\isacharcolon}\ {\isachardoublequoteopen}i\ {\isasymRightarrow}\ i{\isachardoublequoteclose}\ \isakeyword{where}\isanewline
\ \ {\isachardoublequoteopen}edrel{\isacharparenleft}A{\isacharparenright}\ {\isacharequal}{\isacharequal}\ {\isacharbraceleft}{\isacharless}x{\isacharcomma}y{\isachargreater}\ {\isasymin}\ A{\isacharasterisk}A\ {\isachardot}\ x\ {\isasymin}\ domain{\isacharparenleft}y{\isacharparenright}{\isacharbraceright}{\isachardoublequoteclose}
\end{isabelle}
Since \isa{edrel(A)} is a subset of a  well-founded relation (the
transitive closure of the membership relation restricted to \isa{A}),
then it is well-founded as well.

\begin{isabelle}
\isacommand{lemma}\isamarkupfalse%
\ wf{\isacharunderscore}edrel\ {\isacharcolon}\ {\isachardoublequoteopen}wf{\isacharparenleft}edrel{\isacharparenleft}A{\isacharparenright}{\isacharparenright}{\isachardoublequoteclose}\isanewline
\ \ \isacommand{apply}\isamarkupfalse%
\ {\isacharparenleft}rule\ wf{\isacharunderscore}subset\ {\isacharbrackleft}of\ {\isachardoublequoteopen}trancl{\isacharparenleft}Memrel{\isacharparenleft}eclose{\isacharparenleft}A{\isacharparenright}{\isacharparenright}{\isacharparenright}{\isachardoublequoteclose}{\isacharbrackright}{\isacharparenright}\isanewline
\ \ \isacommand{apply}\isamarkupfalse%
\ {\isacharparenleft}auto\ simp\ add{\isacharcolon}edrel{\isacharunderscore}sub{\isacharunderscore}memrel\ wf{\isacharunderscore}trancl\ wf{\isacharunderscore}Memrel{\isacharparenright}\isanewline
\ \ \isacommand{done}\isamarkupfalse%
\end{isabelle}
All but one lemma used in the above proof
(\isa{wf{\isacharunderscore}subset},
\isa{wf{\isacharunderscore}trancl}, 
\isa{wf{\isacharunderscore}Memrel}) are already
present in Isabelle/ZF. The remaining technical result has
been proved using the Isar language:
\begin{isabelle}
\isacommand{lemma}\isamarkupfalse%
\ edrel{\isacharunderscore}sub{\isacharunderscore}memrel{\isacharcolon}\ {\isachardoublequoteopen}edrel{\isacharparenleft}A{\isacharparenright}\ {\isasymsubseteq}\ trancl{\isacharparenleft}Memrel{\isacharparenleft}eclose{\isacharparenleft}A{\isacharparenright}{\isacharparenright}{\isacharparenright}{\isachardoublequoteclose}
\end{isabelle}

The formalization of the functional $\mathit{Hv}$ is straightforward and $\val$ is defined using \isa{wfrec}:
\begin{isabelle}
\isacommand{definition}\isamarkupfalse%
\isanewline
\ \ Hv\ {\isacharcolon}{\isacharcolon}\ {\isachardoublequoteopen}i{\isasymRightarrow}i{\isasymRightarrow}i{\isasymRightarrow}i{\isachardoublequoteclose}\ \isakeyword{where}\isanewline
\ \ {\isachardoublequoteopen}Hv{\isacharparenleft}G{\isacharcomma}y{\isacharcomma}f{\isacharparenright}\ {\isacharequal}{\isacharequal}\ {\isacharbraceleft}\ f{\isacharbackquote}x\ {\isachardot}{\isachardot}\ x{\isasymin}\ domain{\isacharparenleft}y{\isacharparenright}{\isacharcomma}\ {\isasymexists}p{\isasymin}P{\isachardot}\ {\isacharless}x{\isacharcomma}p{\isachargreater}\ {\isasymin}\ y\ {\isasymand}\ p\ {\isasymin}\ G\ {\isacharbraceright}{\isachardoublequoteclose}\isanewline
\isanewline
\isacommand{definition}\isamarkupfalse%
\isanewline
\ \ val\ {\isacharcolon}{\isacharcolon}\ {\isachardoublequoteopen}i{\isasymRightarrow}i{\isasymRightarrow}i{\isachardoublequoteclose}\ \isakeyword{where}\isanewline
\ \ {\isachardoublequoteopen}val{\isacharparenleft}G{\isacharcomma}{\isasymtau}{\isacharparenright}\ {\isacharequal}{\isacharequal}\ wfrec{\isacharparenleft}edrel{\isacharparenleft}eclose{\isacharparenleft}M{\isacharparenright}{\isacharparenright}{\isacharcomma}\ {\isasymtau}{\isacharcomma}\ Hv{\isacharparenleft}G{\isacharparenright}{\isacharparenright}{\isachardoublequoteclose}
\end{isabelle}
Then we can recover the recursive expression~(\ref{eq:def-val}) thanks to the
following lemma:
\begin{isabelle}
\isacommand{lemma}\isamarkupfalse%
\ def{\isacharunderscore}val{\isacharcolon}
\isanewline
   {\isachardoublequoteopen}x{\isasymin}M\ {\isasymLongrightarrow}\ val{\isacharparenleft}G{\isacharcomma}x{\isacharparenright}\ {\isacharequal}\ {\isacharbraceleft}val{\isacharparenleft}G{\isacharcomma}t{\isacharparenright}\ {\isachardot}{\isachardot}\ t{\isasymin}domain{\isacharparenleft}x{\isacharparenright}{\isacharcomma}\ {\isasymexists}p{\isasymin}P\ {\isachardot}\ {\isasymlangle}t{\isacharcomma}\ p{\isasymrangle}{\isasymin}x\ {\isasymand}\ p{\isasymin}G{\isacharbraceright}{\isachardoublequoteclose}
\end{isabelle}

We can finally define the generic extension of $M$ by $G$, also
setting up the notation $M[G]$ for it:
\begin{isabelle}
\isacommand{definition}\isamarkupfalse%
\isanewline
\ \ GenExt\ {\isacharcolon}{\isacharcolon}\ {\isachardoublequoteopen}i{\isasymRightarrow}i{\isachardoublequoteclose}\ {\isacharparenleft}{\isachardoublequoteopen}M{\isacharbrackleft}{\isacharunderscore}{\isacharbrackright}{\isachardoublequoteclose}{\isacharparenright}\ \isakeyword{where} \isanewline
\ \ {\isachardoublequoteopen}GenExt{\isacharparenleft}G{\isacharparenright}{\isacharequal}{\isacharequal}\ {\isacharbraceleft}val{\isacharparenleft}G{\isacharcomma}{\isasymtau}{\isacharparenright}{\isachardot}\ {\isasymtau}\ {\isasymin}\ M{\isacharbraceright}{\isachardoublequoteclose}
\end{isabelle}
It is conventional in Isabelle/ZF to define introduction and
destruction rules for definitions like \isa{GenExt}; in our case, it is
enough to know $x\in M$ in order to know $\val(G,x) \in M[G]$:
\begin{isabelle}
\isacommand{lemma}\isamarkupfalse%
\ GenExtI{\isacharcolon}\ \ {\isachardoublequoteopen}x\ {\isasymin}\ M\ {\isasymLongrightarrow}\ val{\isacharparenleft}G{\isacharcomma}x{\isacharparenright}\ {\isasymin}\ M{\isacharbrackleft}G{\isacharbrackright}{\isachardoublequoteclose}
\end{isabelle}
\noindent The destruction rule corresponding to the generic extension says
that any $x \in M[G]$ comes from some $\tau \in M$ via $\val$. 
\begin{isabelle}
\isacommand{lemma}\isamarkupfalse%
\ GenExtD{\isacharcolon}\ \ {\isachardoublequoteopen}x\ {\isasymin}\ M{\isacharbrackleft}G{\isacharbrackright}\ {\isasymLongrightarrow}\ {\isasymexists}{\isasymtau}{\isasymin}M{\isachardot}\ x\ {\isacharequal}\ val{\isacharparenleft}G{\isacharcomma}{\isasymtau}{\isacharparenright}{\isachardoublequoteclose}
\end{isabelle}


We now provide names for elements in $M$. That is, for each $x\in M$,
we define $\chk(x)$ (usually denoted by $\check{x}$ in the literature)
such that $\val(G,\chk(x))=x$. This will show that $M\sbq M[G]$, with
a caveat we make explicit in the end of this section. As explained in
the introduction, the fact that $M[G]$ extends $M$ is crucial to show
that $\ZFC$ holds in the former.
The definition of $\chk(x)$ is a straightforward $\in$-recursion:
\begin{equation}
  \label{eq:def-check}
  \chk(x)\defi\{\lb\chk(y),\1\rb : y\in x\}
\end{equation}
Now the set-relation argument for \isa{wfrec} is the membership
relation restricted to a set \isa{A}, \isa{Memrel(A)}.

\begin{isabelle}
\isacommand{definition}\isamarkupfalse%
\ \isanewline
\ \ Hcheck\ {\isacharcolon}{\isacharcolon}\ {\isachardoublequoteopen}{\isacharbrackleft}i{\isacharcomma}i{\isacharbrackright}\ {\isasymRightarrow}\ i{\isachardoublequoteclose}\ \isakeyword{where}\isanewline
\ \ {\isachardoublequoteopen}Hcheck{\isacharparenleft}z{\isacharcomma}f{\isacharparenright}\ \ {\isacharequal}{\isacharequal}\ {\isacharbraceleft}\ {\isacharless}f{\isacharbackquote}y{\isacharcomma}one{\isachargreater}\ {\isachardot}\ y\ {\isasymin}\ z{\isacharbraceright}{\isachardoublequoteclose}\isanewline
\isanewline
\isacommand{definition}\isamarkupfalse%
\isanewline
\ \ check\ {\isacharcolon}{\isacharcolon}\ {\isachardoublequoteopen}i\ {\isasymRightarrow}\ i{\isachardoublequoteclose}\ \isakeyword{where}\isanewline
\ \ {\isachardoublequoteopen}check{\isacharparenleft}x{\isacharparenright}\ {\isacharequal}{\isacharequal}\ wfrec{\isacharparenleft}Memrel{\isacharparenleft}eclose{\isacharparenleft}{\isacharbraceleft}x{\isacharbraceright}{\isacharparenright}{\isacharparenright}{\isacharcomma}\ x\ {\isacharcomma}\ Hcheck{\isacharparenright}{\isachardoublequoteclose}
\end{isabelle}
Here, \isa{eclose} returns the (downward) $\in$-closure of its
argument. The main lemmas about $\val$ and $\chk$ require some
instances of replacement for $M$; we set up a locale to assemble these
assumptions:
\begin{isabelle}
\isacommand{locale}\isamarkupfalse%
\ M{\isacharunderscore}extra{\isacharunderscore}assms\ {\isacharequal}\ forcing{\isacharunderscore}data\ {\isacharplus}\isanewline
\ \ \isakeyword{assumes}\ check{\isacharunderscore}in{\isacharunderscore}M\ {\isacharcolon}\ \ \  {\isachardoublequoteopen}{\isasymAnd}x{\isachardot}\ x\ {\isasymin}\ M\ {\isasymLongrightarrow}\ check{\isacharparenleft}x{\isacharparenright}\ {\isasymin}\ M{\isachardoublequoteclose}\isanewline
\ \ \ \isakeyword{and}\ sats{\isacharunderscore}upair{\isacharunderscore}ax{\isacharcolon}\ \ \ \ {\isachardoublequoteopen}upair{\isacharunderscore}ax{\isacharparenleft}{\isacharhash}{\isacharhash}M{\isacharparenright}{\isachardoublequoteclose}\isanewline
\ \ \ \isakeyword{and}\ repl{\isacharunderscore}check{\isacharunderscore}pair{\isacharcolon}\ {\isachardoublequoteopen}strong{\isacharunderscore}replacement{\isacharparenleft}{\isacharhash}{\isacharhash}M{\isacharcomma}{\isasymlambda}p\ y{\isachardot}\ y\ {\isacharequal}{\isacharless}check{\isacharparenleft}p{\isacharparenright}{\isacharcomma}p{\isachargreater}{\isacharparenright}{\isachardoublequoteclose}
\end{isabelle}

The first assumption asserts that all the relevant names are indeed in
$M$ (i.e., $\chk(x) \in M$ if $x\in M$) and it is needed to prove that
$\val(G,\chk(x))=x$. It will take a serious effort to fulfill this
assumption: One of the hardest parts of Paulson's formalization of
constructibility involves showing that models are closed under
recursive construction. We will eventually formalize that if
$M\models\ZFC$ and the arguments of \isa{wfrec} are in $M$, then its
value also is. This will require to adapt to ctm models several
locales defined in \cite{paulson_2003} that were intended to be used
for the class of constructible sets. Notice that the only requirement
on the set \isa{G} is that it contains the top element of the poset
\isa{P}.

\begin{isabelle}
\isacommand{lemma}\isamarkupfalse%
\ valcheck\ {\isacharcolon}\ \isanewline
\ \ \isakeyword{assumes}\ {\isachardoublequoteopen}one\ {\isasymin}\ G{\isachardoublequoteclose}\isanewline
\ \ \isakeyword{shows}\ {\isachardoublequoteopen}y\ {\isasymin}\ M\ {\isasymLongrightarrow}\ val{\isacharparenleft}G{\isacharcomma}check{\isacharparenleft}y{\isacharparenright}{\isacharparenright}\ \ {\isacharequal}\ y{\isachardoublequoteclose}
\end{isabelle}

\subsection{Basic results about the generic extension}

We turn now to prove that $M[G]$ is transitive and $G\in
M[G]$. Showing that $M[G]$ is transitive amounts to prove $y \in M[G]$
for any $x \in M[G]$ and $y \in x$. 

\begin{isabelle}
\isacommand{lemma}\isamarkupfalse%
\ trans{\isacharunderscore}Gen{\isacharunderscore}Ext{\isacharprime}\ {\isacharcolon}\isanewline
\ \ \isakeyword{assumes}\ \ {\isachardoublequoteopen}x\ {\isasymin}\ M{\isacharbrackleft}G{\isacharbrackright}{\isachardoublequoteclose}\ \ \isakeyword{and}\ \ {\isachardoublequoteopen}y\ {\isasymin}\ x{\isachardoublequoteclose}\ \isanewline
\ \ \isakeyword{shows}\ \ \ \ {\isachardoublequoteopen}y\ {\isasymin}\ M{\isacharbrackleft}G{\isacharbrackright}{\isachardoublequoteclose}
\end{isabelle}
The proof of this lemma is straightforward because from $x \in M[G]$
we can obtain $\tau \in M$ such that $x = \val(G,\tau)$. Notice also
that using the characterization of $\val$ given by
\isa{def{\isacharunderscore}val} we can extract some
$\theta \in \dom(\tau)$ such that $y =\val(G,\theta)$; to conclude
$\val(G,\theta) \in M[G]$ it is enough to prove $\theta \in M$, which
follows from the transitivity of $M$.

In contrast, the proof that $G\in M[G]$ is more demanding. In fact, we
set $\dot{G} = \{ \langle \check{p},p\rangle \,|\, p \in P \}$ as a
putative name for $G$. Proving that $\dot{G}$ is in fact a name for
$G$ requires to prove that $\dot{G} \in M$, using an instance of
replacement for $M$ (\noindent namely that given by the assumption
\isa{repl{\isacharunderscore}check{\isacharunderscore}pair}), and
then proving that $\val(G,\dot{G})=G$.

\begin{isabelle}
\isacommand{definition}\isamarkupfalse%
\isanewline
\ \ G{\isacharunderscore}dot\ {\isacharcolon}{\isacharcolon}\ {\isachardoublequoteopen}i{\isachardoublequoteclose}\ \isakeyword{where}\isanewline
\ \ {\isachardoublequoteopen}G{\isacharunderscore}dot\ {\isacharequal}{\isacharequal}\ {\isacharbraceleft}{\isacharless}check{\isacharparenleft}p{\isacharparenright}{\isacharcomma}p{\isachargreater}\ {\isachardot}\ p{\isasymin}P{\isacharbraceright}{\isachardoublequoteclose}
\end{isabelle}

\begin{isabelle}
\isacommand{lemma}\isamarkupfalse%
\ G{\isacharunderscore}dot{\isacharunderscore}in{\isacharunderscore}M\ {\isacharcolon} {\isachardoublequoteopen}G{\isacharunderscore}dot\ {\isasymin}\ M{\isachardoublequoteclose}
\isanewline\isanewline
\isacommand{lemma}\isamarkupfalse%
\ val{\isacharunderscore}G{\isacharunderscore}dot\ {\isacharcolon}\isanewline
\ \ \isakeyword{assumes}\ {\isachardoublequoteopen}G\ {\isasymsubseteq}\ P{\isachardoublequoteclose}\ \ \isakeyword{and}\ \ 
{\isachardoublequoteopen}one\ {\isasymin}\ G{\isachardoublequoteclose}\ \isanewline
\ \ \isakeyword{shows}\ {\isachardoublequoteopen}val{\isacharparenleft}G{\isacharcomma}G{\isacharunderscore}dot{\isacharparenright}\ {\isacharequal}\ G{\isachardoublequoteclose}
\end{isabelle}


\section{Pairing in the generic extension}
\label{sec:pairing-generic-extension}

In this section we show that the generic extension satisfies the
pairing axiom; the purpose of this section is to show how to prove
that $M[G]$ models one of the axioms of $\ZFC$, assuming that $M$
satisfies $\ZFC$.\footnote{The proof that $M[G]$ satisfies pairing
  only needs that $M$ satisfies pairing.} In the locale
\isa{M{\isacharunderscore}extra{\isacharunderscore}assms} we stated
the assumption \isa{sats{\isacharunderscore}upair{\isacharunderscore}ax} which captures that
$M$ satisfies pairing. We use \emph{relativized} versions of the axioms
in order to express satisfaction.


As we have already mentioned, in Paulson's library, the relativized
versions of the $\ZFC$ axioms are defined for classes (which are
defined as predicates over sets). The definition
\isa{upair{\isacharunderscore}ax} corresponds to the Pairing Axiom:

\begin{isabelle}
\isacommand{definition}\isamarkupfalse%
\isanewline
\ \ upair\ {\isacharcolon}{\isacharcolon}\ {\isachardoublequoteopen}{\isacharbrackleft}i{\isasymRightarrow}o{\isacharcomma}i{\isacharcomma}i{\isacharcomma}i{\isacharbrackright}\ {\isasymRightarrow}\ o{\isachardoublequoteclose}\ \isakeyword{where}\isanewline
 {\isachardoublequoteopen}upair{\isacharparenleft}C{\isacharcomma}a{\isacharcomma}b{\isacharcomma}z{\isacharparenright}\ {\isacharequal}{\isacharequal}\ a\ {\isasymin}\ z\ {\isasymand}\ b\ {\isasymin}\ z\ {\isasymand}\ {\isacharparenleft}{\isasymforall}x{\isacharbrackleft}C{\isacharbrackright}{\isachardot}\ x{\isasymin}z\ {\isasymlongrightarrow}\ x\ {\isacharequal}\ a\ {\isasymor}\ x\ {\isacharequal}\ b{\isacharparenright}{\isachardoublequoteclose}
\end{isabelle}
%
%
\begin{isabelle}
\isacommand{definition}\isamarkupfalse%
\isanewline
\ \ upair{\isacharunderscore}ax\ {\isacharcolon}{\isacharcolon}\ {\isachardoublequoteopen}{\isacharparenleft}i{\isasymRightarrow}o{\isacharparenright}\ {\isasymRightarrow}\ o{\isachardoublequoteclose}\ \isakeyword{where}\isanewline
 {\isachardoublequoteopen}upair{\isacharunderscore}ax{\isacharparenleft}C{\isacharparenright}\ {\isacharequal}{\isacharequal}\ {\isasymforall}x{\isacharbrackleft}C{\isacharbrackright}{\isachardot}\ {\isasymforall}y{\isacharbrackleft}C{\isacharbrackright}{\isachardot}\ {\isasymexists}z{\isacharbrackleft}C{\isacharbrackright}{\isachardot}\ upair{\isacharparenleft}C{\isacharcomma}x{\isacharcomma}y{\isacharcomma}z{\isacharparenright}{\isachardoublequoteclose}
\end{isabelle}



We state the main result of this section in the context
\isa{M{\isacharunderscore}extra{\isacharunderscore}assms}.
\begin{isabelle}
\isacommand{lemma}\isamarkupfalse%
\ pairing{\isacharunderscore}axiom\ {\isacharcolon}\ \isanewline
\ \ {\isachardoublequoteopen}one\ {\isasymin}\ G\ {\isasymLongrightarrow}\ upair{\isacharunderscore}ax{\isacharparenleft}{\isacharhash}{\isacharhash}M{\isacharbrackleft}G{\isacharbrackright}{\isacharparenright}{\isachardoublequoteclose}
\end{isabelle}

Let $x$ and $y$ be elements in $M[G]$. By definition of the generic extension, there exist
elements $\tau$ and $\rho$ in $M$ such that $x = \val(G,\tau)$ and
$y = \val(G,\rho)$.  We need to find an element in $M[G]$ that contains exactly
these elements; for that we should construct a name $\sigma\in M$ such that
$\val(G,\sigma) = \{ \val(G,\tau) , \val(G,\rho) \}$. 

The candidate, motivated by the definition of $\chk$,  is
$\sigma = \{\langle \tau , \mathrm{one} \rangle , \langle \rho , \mathrm{one} \rangle \}$. 
Our remaining tasks are to show 
\begin{enumerate}
  \item \label{item:1}$\sigma \in M$, and
  \item \label{item:2} $\val(G,\sigma) = \{ \val(G,\tau) , \val(G,\rho) \}$
\end{enumerate}

By the implementation of pairs  in $\ZFC$, showing (\ref{item:1})
involves using that the
pairing axiom holds in $M$ and the absoluteness of pairing
thanks to $M$ being transitive. 

\begin{isabelle}
\isacommand{lemma}\isamarkupfalse%
\ pairs{\isacharunderscore}in{\isacharunderscore}M\ {\isacharcolon}\ \isanewline
\ \ {\isachardoublequoteopen}\ {\isasymlbrakk}\ a\ {\isasymin}\ M\ {\isacharsemicolon}\ b\ {\isasymin}\ M\ {\isacharsemicolon}\ c\ {\isasymin}\ M\ {\isacharsemicolon}\ d\ {\isasymin}\ M\ {\isasymrbrakk}\ {\isasymLongrightarrow}\ {\isacharbraceleft}{\isasymlangle}a{\isacharcomma}c{\isasymrangle}{\isacharcomma}{\isasymlangle}b{\isacharcomma}d{\isasymrangle}{\isacharbraceright}\ {\isasymin}\ M{\isachardoublequoteclose}
\end{isabelle}

Item (\ref{item:1}) then follows because \isa{\isasymtau}, \isa{\isasymrho} and
\isa{one} belong to \isa{M} (the last fact holds because \isa{one\isasymin P}, \isa{P\isasymin M} and
\isa{M} is transitive).

\begin{isabelle}
\isacommand{lemma}\isamarkupfalse%
\ sigma{\isacharunderscore}in{\isacharunderscore}M\ {\isacharcolon}\isanewline
\ \ {\isachardoublequoteopen} one\ {\isasymin}\ G\ {\isasymLongrightarrow}\ {\isasymtau}\ {\isasymin}\ M\ {\isasymLongrightarrow}\ {\isasymrho}\ {\isasymin}\ M\ {\isasymLongrightarrow}\ {\isacharbraceleft}{\isasymlangle}{\isasymtau}{\isacharcomma}one{\isasymrangle}{\isacharcomma}{\isasymlangle}{\isasymrho}{\isacharcomma}one{\isasymrangle}{\isacharbraceright}\ {\isasymin}\ M{\isachardoublequoteclose}

\isacommand{by}\isamarkupfalse%
\ {\isacharparenleft}rule\ pairs{\isacharunderscore}in{\isacharunderscore}M{\isacharcomma}simp{\isacharunderscore}all\ add{\isacharcolon}\ upair{\isacharunderscore}ax{\isacharunderscore}def\ one{\isacharunderscore}in{\isacharunderscore}M{\isacharparenright}%
\end{isabelle}

Under the assumption that \isa{one} belongs to the set \isa{G},
(\ref{item:2}) follows from \isa{def\_val} almost automatically:

\begin{isabelle}
\isacommand{lemma}\isamarkupfalse%
\ valsigma\ {\isacharcolon}\isanewline
\ \ {\isachardoublequoteopen}one\ {\isasymin}\ G\ {\isasymLongrightarrow}\ {\isacharbraceleft}{\isasymlangle}{\isasymtau}{\isacharcomma}one{\isasymrangle}{\isacharcomma}{\isasymlangle}{\isasymrho}{\isacharcomma}one{\isasymrangle}{\isacharbraceright}\ {\isasymin}\ M\ {\isasymLongrightarrow}\isanewline
\ \ \ val{\isacharparenleft}G{\isacharcomma}{\isacharbraceleft}{\isasymlangle}{\isasymtau}{\isacharcomma}one{\isasymrangle}{\isacharcomma}{\isasymlangle}{\isasymrho}{\isacharcomma}one{\isasymrangle}{\isacharbraceright}{\isacharparenright}\ {\isacharequal}\ {\isacharbraceleft}val{\isacharparenleft}G{\isacharcomma}{\isasymtau}{\isacharparenright}{\isacharcomma}val{\isacharparenleft}G{\isacharcomma}{\isasymrho}{\isacharparenright}{\isacharbraceright}{\isachardoublequoteclose}
\end{isabelle}

\section{Conclusions and future work}
There are several technical milestones that have to be reached in the
course of a formalization of the theory of forcing. The first one, and most
obvious, is the bulk of set- and meta-theoretical concepts needed to work
with. This pushed us, in a sense,  into building on top of Isabelle/ZF,
since we know of no other development in set theory of such
depth (and breadth). In this paper we worked on setting the stage for the work with
generic extensions; in particular, this involves some purely mathematical
results, as the Rasiowa-Sikorski lemma. 

Other milestones in this formalization project
involve 
\begin{enumerate}
\item the definition
  of the forcing relation, 
\item proving the Fundamental Theorem of forcing
  (that relates truth in $M$ to that in $M[G]$), and 
\item using it to show
  that $M[G]\models \ZFC$. 
\end{enumerate}
The theory is very modular and this is
witnessed by the fact 
that the last goal does not depend on the proof of the Fundamental
Theorem nor on the definition of the forcing relation. Our next task
will be to obtain the last goal in that enumeration. 

To this end, we will develop an interface between Paulson's
relativization results and countable models of $\ZFC$. This will show
that every ctm $M$ is closed under well-founded recursion and, in
particular, that contains names for each of its
elements. Consequently, the proof of  $M\sbq M[G]$ will be
complete. A landmark will be to prove the Axiom Scheme
of Separation (the first that needs to use the machinery of forcing
nontrivially). As a part of the new formalization, we will provide
Isar versions of the longer applicative proofs presented in this work.

\ack{We'd like to thank the anonymous referees for reading the paper
  carefully and for their detailed and constructive criticism.}

\providecommand{\noopsort}[1]{}
\begin{small}\end{small}

\end{document}